\documentstyle[aps,eqsecnum,preprint,tighten]{revtex}

\draft
\begin{document}
\title{Reflectionless tunneling through a double-barrier NS junction}
\author{J. A. Melsen and C. W. J. Beenakker}
\address{Instituut-Lorentz, University of Leiden\\
P.O. Box 9506, 2300 RA Leiden, The Netherlands}
\maketitle

\narrowtext
\begin{abstract}
The resistance is computed of an ${\rm NI}_{1}{\rm NI}_{2}{\rm S}$ junction,
where N = normal metal, S = superconductor, and ${\rm I}_{i}$ = insulator or
tunnel barrier (transmission probability per mode $\Gamma_{i}$). The ballistic
case is considered, as well as the case that the region between the two
barriers contains disorder (mean free path $l$, barrier separation $L$). It is
found that the resistance at fixed $\Gamma_{2}$ shows a {\em minimum\/} as a
function of $\Gamma_{1}$, when $\Gamma_{1}\approx\sqrt{2}\Gamma_{2}$,
provided $l\gtrsim\Gamma_2 L$. The minimum is explained in terms of the
appearance of transmission eigenvalues close to one, analogous to the
``reflectionless tunneling'' through a NIS junction with a disordered normal
region. The theory is supported by numerical simulations.
\end{abstract}
\bigskip

\section{Introduction}
Reflectionless tunneling is a novel quantum interference effect which occurs
when dissipative normal current is converted into dissipationless supercurrent
at the interface between a normal metal (N) and a superconductor (S)
\cite{Houches}. Experimentally, the effect is observed as a peak in the
differential conductance around zero voltage or around zero magnetic field
\cite{Kas91}. Its name refers to the fact that, for full phase coherence, the
Andreev-reflected quasiparticle can tunnel through the potential barrier at the
NS interface without suffering reflections. (The potential barrier can be the
insulator (I) in an NIS junction, or the Schottky barrier in a
semiconductor--superconductor junction.) Application of a voltage or magnetic
field destroys the phase coherence between electrons and holes, and thus
reduces the conductance of the junction. We now have a good theoretical
understanding of the effect, based on a combination of numerical
\cite{Mar93,Tak93a}, and analytical work
\cite{Wee92,Tak92c,Vol93,Hek93,Bee94b,Naz94}. The basic requirement for
reflectionless tunneling is that the normal region has a resistance which is
larger than the resistance of the interface. In that case the disorder is able
to open a fraction of the tunneling channels, {\em i.e.} it induces the
appearance
of transmission eigenvalues close to one \cite{Naz94}. As a result of these
open channels, the resistance has a linear dependence on the transparency of
the interface, instead of the quadratic dependence expected for Andreev
reflection \cite{Shelankov} (which is a two-particle process).

The purpose of this work is to present a study of reflectionless tunneling in
its simplest form, when the resistance of the normal metal is due to a second
tunnel barrier, in series with the barrier at the NS interface. This allows an
exact calculation, which shows many of the features of the more complicated
case when the resistance of the normal region is due to disorder. Furthermore,
the double-barrier geometry provides an experimentally realizable model system,
for example in tunneling from an STM into a superconductor via a metal particle
\cite{Hesl94}.

The outline of this paper is as follows. In Section 2 we consider the problem
of a ${\rm NI}_{1}{\rm NI}_{2}{\rm S}$ junction without disorder. We compute
the resistance of the junction as a function of the transmission probabilities
per mode $\Gamma_{1}$ and $\Gamma_{2}$ of the two barriers. The resistance at
fixed $\Gamma_{2}$ shows a {\em minimum\/} as a function of $\Gamma_{1}$ when
$\Gamma_{1}\approx\sqrt{2}\Gamma_{2}\equiv\Gamma$. The resistance in the
minimum depends
{\em linearly\/} on $1/\Gamma$, in contrast to the quadratic dependence in the
case of a single barrier. In Section 3 we apply a recent scaling theory
\cite{Bee94b}, to find the influence on the resistance minimum of disorder in
the region between the barriers (length $L$, mean free path $l$). The
resistance minimum persists as long as $l\gtrsim\Gamma L$. In the diffusive
regime ($l\ll L$) our results agree with a previous Green's function
calculation by Volkov, Zaitsev, and Klapwijk \cite{Vol93}. The analytical
results are supported by numerical simulations, using the recursive Green's
function technique~\cite{Bar91}. We conclude in Section 4.

\section{NINIS junction without disorder}

We consider a ${\rm NI_1NI_2S}$ junction, where ${\rm N}$ = normal metal, ${\rm
S}$ = superconductor, and ${\rm I}_i$ = insulator
or tunnel barrier (see inset of Fig.~1).
The transmission probability per mode of ${\rm I}_i$ is denoted by $\Gamma_i$.
For simplicity we neglect the mode-dependence of $\Gamma_i$.
In this section, we assume ballistic motion between the barriers. (The effect
of disorder in the normal region is considered in Sec. III.) A straightforward
calculation yields the transmission probabilities $T_n$ of the two barriers in
series,
\begin{equation}
T_n = (a + b \cos\phi_n)^{-1},
\label{eq:tnphin}
\end{equation}
where
\begin{mathletters}
\label{eq:ab}
\begin{eqnarray}
a & = & 1+\frac{2-\Gamma_1-\Gamma_2}{\Gamma_1\Gamma_2},\\
b & = & \frac{2\,(1-\Gamma_1)^{1/2}(1-\Gamma_2)^{1/2}}{\Gamma_1\Gamma_2},
\end{eqnarray}
\end{mathletters}%
and $\phi_n$ is the phase accumulated between the barriers by mode
$n=1,2,\cdots N$ (with $N$ the number of propagating modes at the Fermi level).
If we substitute $\Gamma_i = 1/\mbox{cosh}^2 \alpha_i$ ($\alpha_i \geq 0$), the
coefficients $a$ and $b$ can be rewritten as
\begin{mathletters}
\label{eq:abalpha}
\begin{eqnarray}
a & = & \case{1}{2} + \case{1}{2} \cosh 2\alpha_1 \cosh 2\alpha_2, \\
b & = & \case{1}{2} \sinh 2\alpha_1 \sinh 2\alpha_2.
\end{eqnarray}
\end{mathletters}

Since the transmission matrix $t$ is diagonal, the transmission probabilities
$T_n$
are identical to the eigenvalues of $tt^{\dagger}$.
We use the general relationship between the conductance $G_{\rm NS}
\equiv G_{\rm NINIS}$ of the NINIS junction and the transmission eigenvalues of
the normal region~\cite{carlo},
\begin{equation}
G_{\rm NS}= \frac{4 e^2}{h} \sum_{n=1}^N \frac{T_n^2}{(2-T_n)^2},
\label{eq:carlo}
\end{equation}
which is the analogue of the Landauer formula,
\begin{equation}
G_{\rm N}= \frac{2 e^2}{h} \sum_{n=1}^N T_n,
\label{eq:land}
\end{equation}
for the conductance $G_{\rm N} \equiv G_{\rm NININ}$ in the normal state.
We assume that $L \gg \lambda_{\rm F}$ ($\lambda_{\rm F}$ is the Fermi
wavelength) and $N\Gamma_i \gg 1$,
so that the conductance is not dominated by a single resonance.
In this case, the phases $\phi_n$ are distributed uniformly in the
interval $(0,2\pi)$ and we may replace the summations in
Eqs.~(\ref{eq:carlo}),~(\ref{eq:land}) by integrals over $\phi$: $\sum_{n=1}^N
f(\phi_n) \rightarrow (N/2\pi) \int_0^{2\pi} {\rm d}\phi\,f(\phi)$. The result
is
\begin{eqnarray}
G_{\rm NS} & = & \frac{4 e^2 N}{h} \frac{\cosh 2 \alpha_1\cosh 2
\alpha_2}{\left( \cosh^22 \alpha_1 + \cosh^2 2 \alpha_2 - 1
\right)^{3/2}},\label{gnsintphi}\\
G_{\rm N} & = & \frac{4 e^2 N}{h} (\cosh 2 \alpha_1+\cosh 2
\alpha_2)^{-1}\label{gnintphi}.
\end{eqnarray}
These expressions are symmetric in the indices 1 and 2: it does not matter
which of the two barriers is closest to the superconductor.

In the same way we can compute the entire distribution of the transmission
eigenvalues, $\rho(T) \equiv \sum_n \delta(T-T_n) \rightarrow (N/2\pi)
\int_0^{2\pi} {\rm d}\phi\,\delta(T-T(\phi))$. Substituting $T(\phi) =
(a+b\cos\phi)^{-1}$ from Eq.~(\ref{eq:tnphin}), we find
\begin{equation}
\rho(T) = \frac{N}{\pi T}\left(b^2T^2 -(aT-1)^2\right)^{-1/2}.
\label{eq:rhot}
\end{equation}

In Fig.~1 we plot the resistance $R_{\rm N} = 1/G_{\rm N}$ and $R_{\rm NS} =
1/G_{\rm NS}$, following from
Eqs.~(\ref{gnsintphi}) and~(\ref{gnintphi}). Notice
that $R_{\rm N}$ follows Ohm's law,
\begin{equation}
R_{\rm N} = \frac{h}{2N e^2}(1/\Gamma_1 + 1/\Gamma_2 -1),
\end{equation}
as expected from classical considerations.
In contrast, the resistance $R_{\rm NS}$ has a {\em minimum} if one of the
$\Gamma$'s is varied while keeping the other fixed. This resistance minimum
cannot be explained by classical series addition of barrier resistances. If
$\Gamma_2 \ll 1$ is fixed and $\Gamma_1$ is varied, as in Fig.~1, the minimum
occurs when $\Gamma_1 = \sqrt{2} \Gamma_2$. The minimal resistance $R_{\rm
NS}^{\rm min}$ is of the same order of magnitude as the resistance $R_{\rm N}$
in the normal state at the same value of $\Gamma_1$ and $\Gamma_2$. (For
$\Gamma_2 \ll 1$, $R_{\rm NS}^{\rm min} = 1.52\, R_{\rm N}$) In particular, we
find that $R_{\rm NS}^{\rm min}$ depends linearly on $1/\Gamma_i$, whereas for
a single barrier $R_{\rm NS} \propto 1/\Gamma^2$.

The linear dependence on the barrier transparency shows the qualitative
similarity of a ballistic NINIS junction to a disordered NIS junction.
To illustrate the similarity, we compare in Fig.~2 the densities of
transmission eigenvalues throught the normal region. The left panel is for an
NIS junction (computed using the results of Ref.~\cite{Bee94b}), the right
panel is for an NINIS junction (computed from Eq.~(\ref{eq:rhot})). In the NIS
junction, disorder leads to a bimodal distribution $\rho(T)$,
with a peak near zero transmission and another peak near unit transmisssion
(dashed curve). A similar bimodal distribution appears in the ballistic NINIS
junction, for approximately equal transmission probabilities of the two
barriers. There are also differences between the two cases: The NIS junction
has a unimodal $\rho(T)$ if $L/l < 1/\Gamma$, while the NINIS junction has a
bimodal $\rho(T)$ for any ratio of $\Gamma_1$ and $\Gamma_2$. In both cases,
the opening of tunneling channels, {\em i.e.}, the appearance of a peak in
$\rho(T)$ near $T=1$, is the origin for the $1/\Gamma$ dependence of the
resistance.


\section{Effects of disorder}

Let us now investigate what happens to the resistance minimum if the region of
length $L$ between the tunnel barriers contains impurities, with elastic
mean free path $l$. We denote $s \equiv L/l$.
When introducing disorder, it is necessary to consider ensemble-averaged
quantities.
To calculate the ensemble-averaged conductance $\langle G_{\rm NS}\rangle$, we
need
to know the density $\rho$ of the transmission eigenvalues $T_n$ as a function
of $s$.
It is convenient to work with the parameterization
\begin{equation}
T_n = 1/\cosh^2 x_n, \ \ \ x_n \geq 0.
\label{eq:tx}
\end{equation}
The density of the $x_n$'s is defined by
$\rho(x,s) \equiv \langle\sum_n \delta(x-x_n)\rangle$. From
Eq.~(\ref{eq:tnphin}) we know that, for $s = 0$ (no disorder),
\begin{eqnarray}
\rho(x,0) & = & N \int_0^{2\pi} \frac{\mbox{d}\phi}{2\pi}\,
\delta(x-\mbox{arccosh}\sqrt{a + b\cos\phi})\nonumber\\
 & = & \frac{N}{\pi}\sinh 2x\left( b^2 - (a-\cosh^2x)^2\right)^{-\frac{1}{2}},
\label{eq:rhox0}
\end{eqnarray}
for $\mbox{arccosh}\sqrt{a-b} \equiv x_{\rm min} \leq x \leq x_{\rm max} \equiv
\mbox{arccosh}\sqrt{a+b}$.

For $s > 0$ we obtain the density $\rho(x,s)$ from the integro-differential
equation~\cite{mellopichard}
\begin{equation}
\frac{\partial}{\partial s} \rho(x,s) = - \frac{1}{2N}\frac{\partial}{\partial
x} \rho(x,s)\frac{\partial}{\partial x} \int_0^{\infty} \mbox{d}x' \rho(x',s)
\ln | \sinh^2x - \sinh^2x' |,
\label{eq:dmpk}
\end{equation}
which is the large $N$-limit of the scaling equation due to
Dorokhov~\cite{dorokhov} and Mello, Pereyra, and Kumar~\cite{mpk}.
This equation describes the evolution of $\rho(x,s)$ when
an infinitesimal slice of disordered material is added.
With initial condition (\ref{eq:rhox0}) it therefore describes a geometry where
all disorder is on one side of the two tunnel barriers, rather than in between.
In fact, only the total length $L$ of the disordered region matters, and not
the location relative to the barriers. The argument is similar to that in
Ref.~\cite{bm}.
The total transfer matrix $M$ of the normal region is a
product of the transfer matrices of its constituents (barriers and disordered
segments):
$M = M_1 M_2 M_3\cdots$. The probability distribution of $M$
is given by the convolution
$p(M) = p_1*p_2*p_3*\cdots$
of the distributions $p_i$ of transfer matrices $M_i$. The convolution is
defined as
\begin{equation}
p_i*p_j(M)=\int \mbox{d}M'\,p_i(MM'^{-1})p_j(M').
\end{equation}
If for all parts $i$ of the system, $p_i(M_i)$ is a function of the eigenvalues
of $M_i^{\vphantom{\dagger}}M_i^{\dagger}$ only, the convolution of the $p_i$
commutes~\cite{bm}.
The distributions $p_i$ are then called isotropic. A disordered segment
(length $L$, width $W$) has an isotropic distribution if $L \gg W$. A planar
tunnel barrier does not mix the modes, so a priori it does not have an
isotropic distribution. However, if the mode-dependence of the transmission
probabilities is neglected (as we do here), it does not make a difference if we
replace its distribution by an isotropic one.
The commutativity of the convolution of isotropic distributions implies that
the location of the tunnel barriers
with respect to the disordered region does not affect $\rho(x,s)$. The systems
in Figs.~3a, b, and c then have identical statistical properties.


Once $\rho(x,s)$ is known, the conductances $\langle G_{\rm NS}\rangle$ and
$\langle G_{\rm N}\rangle$ can be determined from
\begin{eqnarray}
\langle G_{\rm NS} \rangle & = & \frac{4e^2}{h} \int_0^{\infty}\mbox{d}x\,
\frac{\rho(x,s) }{\cosh^22x},\label{eq:gnsav}\\
\langle G_{\rm N} \rangle & = & \frac{2e^2}{h} \int_0^{\infty}\mbox{d}x\,
\frac{\rho(x,s)}{\cosh^2x},
\end{eqnarray}
where we have substituted Eq.~(\ref{eq:tx}) into Eqs.~(\ref{eq:carlo}),
(\ref{eq:land}).
In Ref.~\cite{Bee94b} a general solution to the evolution equation was obtained
for arbitrary initial condition. It was shown that Eq.~(\ref{eq:dmpk}) can be
mapped onto Euler's equation of hydrodynamics
\begin{equation}
\frac{\partial}{\partial s}U(\zeta,s) + U(\zeta,s) \frac{\partial}{\partial
\zeta} U(\zeta,s) = 0,
\label{eq:euler}
\end{equation}
by means of the substitution
\begin{equation}
U(\zeta,s) = \frac{\sinh 2 \zeta}{2 N} \int_0^{\infty} \mbox{d} x'\,
\frac{\rho(x',s)}{\sinh^2\zeta - \sinh^2x'}.
\label{eq:veloc}
\end{equation}
Here, $U \equiv U_x + {\rm i}U_y$ and $\zeta \equiv x + {\rm i}y$.
Eq.~(\ref{eq:euler}) describes the velocity field $U(\zeta,s)$ of a 2D ideal
fluid at constant pressure in the $x$-$y$ plane. Its solution is\footnote{ The
implicit equation (\ref{eq:impl}) has multiple solutions in the entire complex
plane; we need the solution for which both $\zeta$ and $\zeta -sU(\zeta,s)$ lie
in the strip between the lines $y = 0$ and $y = -\pi/2$.}
\begin{equation}
U(\zeta,s) = U_0(\zeta - sU(\zeta,s)),
\label{eq:impl}
\end{equation}
in terms of the initial value $U_0(\zeta) \equiv U(\zeta,0)$. The probability
distribution $\rho(x,s)$ follows from the
velocity field by inversion of Eq.~(\ref{eq:veloc}),
\begin{equation}
\rho(x,s) = \frac{2 N}{\pi} U_y(x-{\rm i}\epsilon,s),
\label{eq:rhoxs}
\end{equation}
where $\epsilon$ is a positive infinitesimal.

In our case, the initial
velocity field [from Eqs.~(\ref{eq:rhox0}) and~(\ref{eq:veloc})] is
\begin{equation}
U_0(\zeta) = -\case{1}{2} \sinh 2 \zeta \left[ (\cosh^2\zeta - a)^2 - b^2
\right]^{-\frac{1}{2}}.
\label{eq:velo0}
\end{equation}
The resulting density (\ref{eq:rhoxs}) is plotted in Fig.~4 for $\Gamma_1 =
\Gamma_2 \equiv \Gamma$ and several disorder strengths.
The region near $x = 0 $ is of importance
for the conductance (since $x$ near zero corresponds to near-unit
transmission). The number $N_{\rm open} \equiv \rho(0,s)$ is an estimate for
the number of transmission eigenvalues close to 1 (so-called ``open
channels''~\cite{imry}). In the absence of disorder, $N_{\rm open}$ is non-zero
only if $\Gamma_1 \approx \Gamma_2$ (then $a-b=1 \Rightarrow x_{\rm min} = 0$).
{}From Eq.~(\ref{eq:rhox0}) we find $N_{\rm open} = N\Gamma/\pi$ for $s=0$ and
$\Gamma_1 = \Gamma_2 \equiv \Gamma \ll 1$. Adding disorder reduces the number
of open channels.
If $\Gamma_1 \neq \Gamma_2$ there are no open channels for $s = 0$ ($x_{\rm
min} > 0$). Disorder then has the effect of increasing $N_{\rm open}$, such
that $N_{\rm open} \approx N/s$ if $(\Gamma_1 + \Gamma_2)s \gg 1$.
The disorder-induced opening of channels was studied in
Refs.~\cite{Bee94b,Naz94} for the case of a single tunnel barrier.

To test our analytical results for the eigenvalue density $\rho(x,s)$, we have
carried
out numerical simulations, similar to those reported in Ref.~\cite{Bee94b}. The
sample was modeled by a tight-binding
Hamiltonian on a square lattice with lattice constant {\em a}. The tunnel
barriers were accounted for by assigning a non-random potential energy $U_{\rm
B} = 2.3 \, E_{\rm F}$ to a single row of sites at both ends of the lattice,
which corresponds
to a mode averaged barrier transparency $\Gamma_1 = \Gamma_2 = 0.18$. The Fermi
energy
was chosen at 1.5\,$u_0$, with $u_0 = \hbar^2/2 m a^2$. Disorder was
introduced by randomly assigning a value between $\pm \case{1}{2}U_{\rm D}$ to
the on-site potential of the lattice points   between the barriers. The
disorder strength
$U_{\rm D}$ was varied between
0 and 1.5\,$u_0$, corresponding to $s$ between 0 and 11.7.
We considered geometries with both a square disordered region ($285\times 285$
sites, $N=119$)
and a rectangular one ($285\times 75$ sites, $N=31$), to test the geometry
dependence
of our results. In Fig.~5, we compare the integrated eigenvalue density
$\nu(x,s) \equiv N^{-1} \int_0^x \mbox{d}x' \rho(x',s)$ with the numerical
results.

The quantity $\nu(x,s)$ follows directly from our simulations, by plotting the
$x_n$'s in ascending order versus $n/N \equiv \nu$. We want to sample
$\nu(x,s)$ at many points along the $x$-axis, so we need $N$ large.
Since the $x_n$'s are self-averaging (fluctuations are of the order of $1/N$),
it is not necessary to average over many samples.
The data shown in Fig.~5 are from a single realization of the impurity
potential.
There is good agreement with the analytical results.
No geometry dependence is observed, which indicates that the restriction $L \gg
W$ of Eq.~(\ref{eq:dmpk}) can be relaxed to a considerable extent.

Using Eqs.~(\ref{eq:gnsav}) and~(\ref{eq:veloc}), the average conductance
$\langle G_{\rm NS}\rangle$ can be directly expressed in terms of the velocity
field,
\begin{equation}
\langle G_{\rm NS}\rangle = \frac{2 N e^2}{h} \lim_{\zeta \rightarrow -{\rm i}
\pi/4} \frac{\partial}{\partial \zeta} U(\zeta,s).
\label{eq:gu}
\end{equation}
For  $\zeta \rightarrow -{\rm i} \pi/4$, $U \rightarrow {\rm i}U_y$, $U_y > 0$.
The implicit solution
(\ref{eq:impl}) then takes the form
\begin{equation}
\phi \sqrt{ (2a + \mbox{sin}\phi - 1)^2 - 4b^2} = 2s \cos\phi,
\label{eq:phi}
\end{equation}
where $\phi \equiv 2sU_y \in [0,\pi/2]$.
We now use that $\frac{\partial}{\partial \zeta}U(\zeta, s)|_{\zeta=-\frac{{\rm
i}\pi}{4}} = -[\frac{\partial}{\partial s} U(-\frac{{\rm i} \pi}{4},s)] /
U(-\frac{{\rm i} \pi}{4},s)$ [see Eq.~(\ref{eq:euler})].
Combining Eqs.~(\ref{eq:gu}) and~(\ref{eq:phi}) we find
\begin{equation}
\langle G_{\rm NS}\rangle = \frac{2 N e^2}{h} ( s + 1/Q)^{-1},
\label{eq:gnsq}
\end{equation}
where the effective tunnel rate $Q$ is given in terms of the angle $\phi$ in
Eq.~(\ref{eq:phi}) by
\begin{equation}
Q = \frac{\phi}{s \cos\phi} \left( \sin\phi + \frac{\phi^2}{4 s^2}
\left[ \sin\phi + (1-2/\Gamma_1)(1-2/\Gamma_2) \right] \right).
\label{eq:q}
\end{equation}
Eqs.~(\ref{eq:phi})--(\ref{eq:q}) completely determine the conductance of a
double-barrier NS-junction
containing disorder.

In Fig.~6, we plot $\langle R_{\rm NS}\rangle$ for several values of the
disorder, keeping
$\Gamma_2 = 0.1$ fixed and varying the transparency of barrier 1.
For weak disorder
($\Gamma_2 s \ll 1$), the resistance minimum is retained, but its location
moves to larger values of $\Gamma_1$. On increasing the disorder, the minimum
becomes
shallower and eventually disappears. In the  regime of strong disorder
($\Gamma_2 s \gg 1$), the resistance behaves nearly Ohmic.

We stress that these results hold for arbitrary $s\equiv L/l$, all the way from
the ballistic into the diffusive regime.
Volkov, Zaitsev, and Klapwijk~\cite{Vol93}  have computed $\langle G_{\rm
NS}\rangle$ in the diffusive limit $s \gg 1$.
In that limit our Eqs.~(\ref{eq:phi}) and (\ref{eq:q}) take the form:
\begin{equation}
\frac{s \cos\phi}{\phi} = \frac{1}{\Gamma_1}
\sqrt{1 - \left( \frac{\phi}{\Gamma_2 s} \right)^2} +
\frac{1}{\Gamma_2}
\sqrt{1 - \left(\frac{\phi}{\Gamma_1 s} \right)^2},
\label{eq:philimit}
\end{equation}
\begin{equation}
\frac{1}{Q} = \sum_{i=1}^2 \frac{1}{\Gamma_i}\left[1-\left(\frac{\phi}{\Gamma_i
s}\right)^2\right]^{-1/2},
\label{eq:qlimit}
\end{equation}
in precise agreement with Ref.~\cite{Vol93}.
Nazarov's circuit theory~\cite{Naz94}, which is equivalent to the Green's
function theory of Ref.~\cite{Vol93}, also leads to this result for $\langle
G_{\rm NS} \rangle$ in the diffusive regime.

Two limiting cases of Eqs.~(\ref{eq:philimit}) and~(\ref{eq:qlimit}) are of
particular interest. For strong barriers, $\Gamma_1, \Gamma_2 \ll 1$, and
strong disorder, $s \gg 1$, one has the two asymptotic formulas
\begin{eqnarray}
\langle G_{\rm NS} \rangle & = &
\frac{2Ne^2}{h}\frac{\Gamma_1^2\Gamma_2^2}{\left(\Gamma_1^2+\Gamma_2^2\right)^{3/2}}, \ \ \ \mbox{if}\ \ \ \Gamma_1,\Gamma_2 \ll 1/s,\label{eq:glimitls}\\
\langle G_{\rm NS} \rangle & = & \frac{2Ne^2}{h}(s+1/\Gamma_1+1/\Gamma_2)^{-1},
\ \ \ \mbox{if}\ \ \ \Gamma_1,\Gamma_2 \gg 1/s.\label{eq:glimitss}
\end{eqnarray}
Eq.~(\ref{eq:glimitls}) coincides with Eq.~(\ref{gnsintphi}) in the limit
$\alpha_1, \alpha_2 \gg 1$ (recall that $\Gamma_i\equiv 1/\cosh^2\alpha_i$).
This shows that the effect of disorder on the resistance minimum can be
neglected as long as the resistance of the junction is dominated by the
barriers. In this case $\langle G_{\rm NS} \rangle $ depends linearly on
$\Gamma_1$ and $\Gamma_2$ only if $\Gamma_1 \approx \Gamma_2$.
Eq.~(\ref{eq:glimitss}) shows that if the disorder dominates, $\langle G_{\rm
NS} \rangle$ has a linear $\Gamma$-dependence regardless of the relative
magnitude of $\Gamma_1$ and $\Gamma_2$.

\section{Conclusions}
In summary, we have derived an expression for the conductance of a ballistic
NINIS junction in the limit $N\Gamma \gg 1$ that the tunnel resistance is much
smaller than $h/e^2$. In this regime the double-barrier junction contains a
large number of
overlapping resonances, so that in the normal state the resistance depends
monotonically on $1/\Gamma$. In contrast,
the NINIS junction shows a resistance minimum when one of the
barrier transparencies is varied while the other is kept fixed. The minimal
resistance (at $\Gamma_1 \simeq\Gamma_2 \equiv \Gamma$)
is proportional to $1/\Gamma$, instead of the $1/\Gamma^2$ dependence expected
for two-particle tunneling into a superconductor.
This is similar to the reflectionless tunneling which occurs in an NIS
junction. Using the results of the ballistic junction, we have described the
transition
to a disordered NINIS junction by means of an evolution equation for the
transmission eigenvalue density~\cite{Bee94b}. We found that the
resistance minimum is unaffected by disorder, as long as $l \gg L/\Gamma$, {\em
i.e.}, as long as the barrier resistance dominates the junction resistance.
As the disorder becomes more dominant, a transition to a monotonic
$\Gamma$-dependence takes place.
In the limit of diffusive motion between the barriers, our results agree with
Ref.~\cite{Vol93}.

Throughout this paper we have assumed zero temperature, zero magnetic field,
and infinitesimal applied voltage. Each of these quantities is capable of
destroying the phase coherence between the electrons and the Andreev-reflected
holes, which is responsible for the resistance minimum. As far as the
temperature $T$ and voltage $V$ are concerned, we require $k_{\rm B}T, eV \ll
\hbar/\tau_{\rm dwell}$ for the appearance of a resistance minimum, where
$\tau_{\rm dwell}$ is the dwell time of an electron in the region between the
two barriers.
For a ballistic NINIS junction we have $\tau_{\rm dwell} \sim L/v_{\rm
F}\Gamma$,
while for a disordered junction $\tau_{\rm dwell} \sim L^2/v_{\rm F}\Gamma l$
is larger by a factor $L/l$. It follows that the condition on temperature and
voltage becomes more restrictive if the disorder increases, even if the
resistance remains dominated by the barriers. As far as the magnetic field $B$
is concerned, we require $B \ll h/eS$ (with $S$ the area of the junction
perpendicular to $B$), if the motion between the barriers is diffusive. For
ballistic motion the trajectories enclose no flux, so no magetic field
dependence is expected.

A possible experiment to verify our results might be
scanning tunneling microscopy of a metal particle on top of a
superconducting substrate~\cite{Hesl94}. The metal-superconductor interface has
a fixed tunnel probability $\Gamma_2$. The probability $\Gamma_1$ for an
electron to tunnel from STM to particle can be controlled by varying the
distance. (Volkov has recently analyzed this geometry in the regime that the
motion from STM to particle is diffusive rather than by
tunneling~\cite{Vol94}.)
Another possibility is to create an NINIS junction using a two-dimensional
electron gas in contact with a superconductor. The tunnel barriers could then
be implemented by means of two gate electrodes. In this way both transparancies
might be tuned independently.

\acknowledgments
This research was motivated by a discussion with D. Est\`{e}ve, which is
gratefully acknowledged. Financial support was provided by the
``Ne\-der\-land\-se or\-ga\-ni\-sa\-tie voor We\-ten\-schap\-pe\-lijk
On\-der\-zoek'' (NWO), by the ``Stich\-ting voor Fun\-da\-men\-teel
On\-der\-zoek der Ma\-te\-rie'' (FOM), and by the ``Human Capital and
Mobility'' programme of the European Community.

\newpage

{\bf Figure 1\\}
Dependence of the resistances $R_{\rm N}$ and $R_{\rm NS}$ of ballistic NININ
and NINIS structures, respectively, on barrier transparancy $\Gamma_1$,
while transparancy $\Gamma_2=0.1$ is kept fixed [computed from
Eqs.~(\protect\ref{gnsintphi}) and~(\protect\ref{gnintphi})]. The inset shows
the NINIS structure considered.

{\bf Figure 2\\}
Density of normal-state transmission eigenvalues for an NS junction with a
potential
barrier at the interface (transmission probability $\Gamma=0.4$). The
left panel (a) shows the disorder-induced opening of tunneling channels
in an NIS junction (solid curve: $s=0.04$; dotted: $s=0.4$; dashed:
$s=5$; where $s\equiv L/l$). The right panel (b) shows the opening of
channels by a second tunnel barrier (transparancy $\Gamma'$) in an
NINIS junction (solid curve: $\Gamma'=0.95$; dotted: $\Gamma'=0.8$;
dashed: $\Gamma'=0.4$). The curves in (a) are computed from
Ref.\ \protect\cite{Bee94b}, the curves in (b) from Eq.\
(\protect\ref{eq:rhot}). Notice the similarity of the dashed curves.

{\bf Figure 3\\}
The systems a,b, and c are statistically equivalent, if the transfer matrices
of each of the two barriers (solid vertical lines) and the disordered regions
(shaded areas, $L_1+L_2=L$ in case b) have isotropic distributions; in that
case, the position of the disorder with
respect to the barriers does not affect the eigenvalue density $\rho(x,s)$.

{\bf Figure 4\\}
Eigenvalue density $\rho(x,s)$ as a function of $x$ (in units of $s\equiv L/l$)
for $\Gamma_1=\Gamma_2 = 0.2$. Curves a,b,c,d, and e are for $s=0.5, 2, 5, 20,
100$, respectively. In the special case of equal tunnel barriers, open channels
exist already in the absence of disorder.

{\bf Figure 5\\}
Comparison between theory and simulation of the integrated eigenvalue density
for $\Gamma_1=\Gamma_2=0.18$. The labels $a,b,c$ indicate, respectively, $s=0,
3,
11.7$. Solid curves are from Eq.~(\protect\ref{eq:impl}); data points are the
$x_n$'s from the simulation plotted in ascending order versus $n/N$. Filled
data
points are for a square geometry, open points are for an aspect ratio
$L/W=3.8$.

{\bf Figure 6\\}
Dependence of the ensemble-averaged resistance $\langle R_{\rm NS}\rangle$ for
a disordered NINIS junction on barrier transparancy $\Gamma_1$, while
$\Gamma_2=0.1$ is kept fixed [computed from Eqs.~(\protect\ref{eq:gnsq})
and~(\protect\ref{eq:q})].
Curves $a,b,c,d$ are for $s=0,2,7,30$, respectively. The resistance minimum
persists for small disorder.

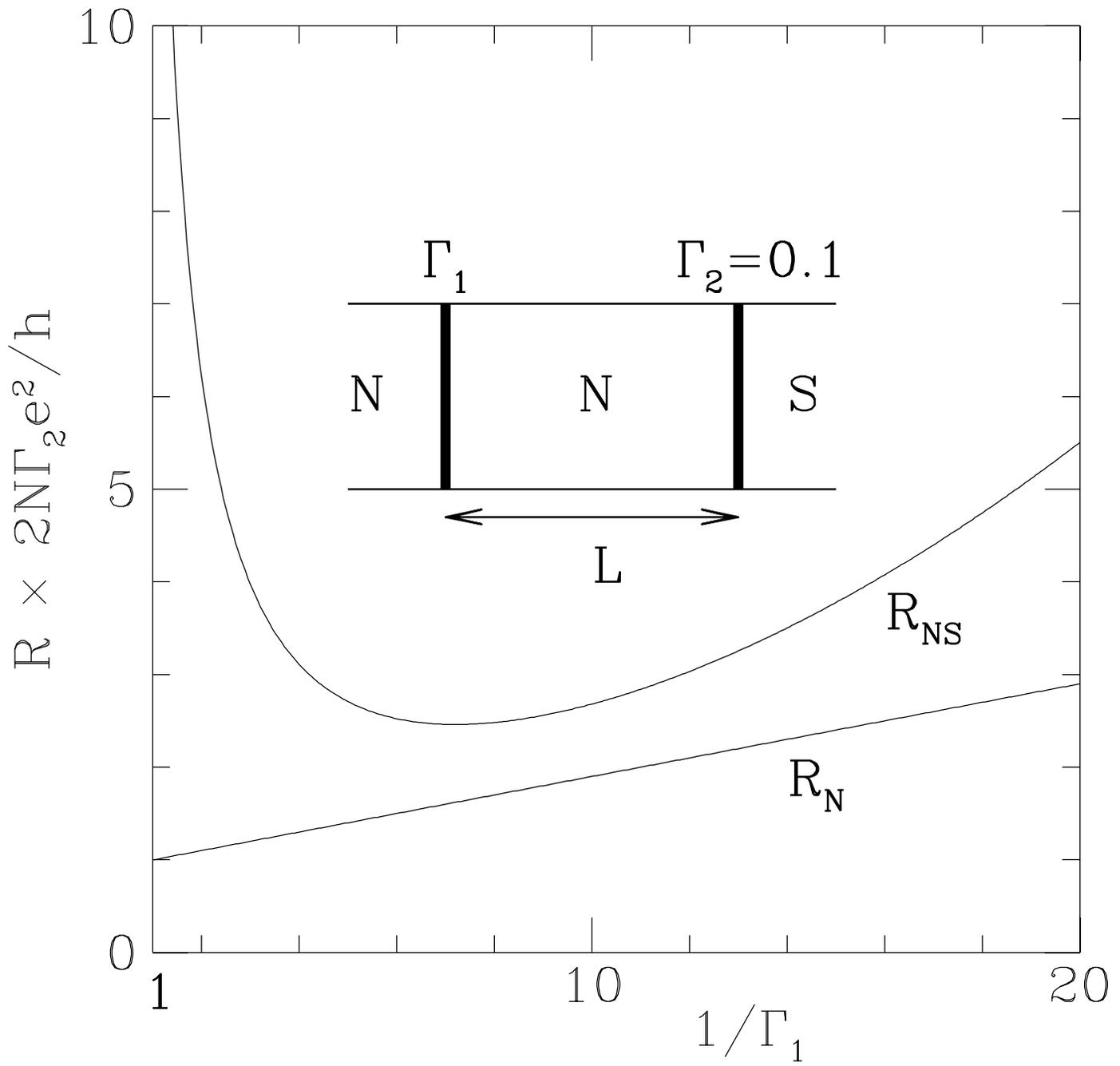
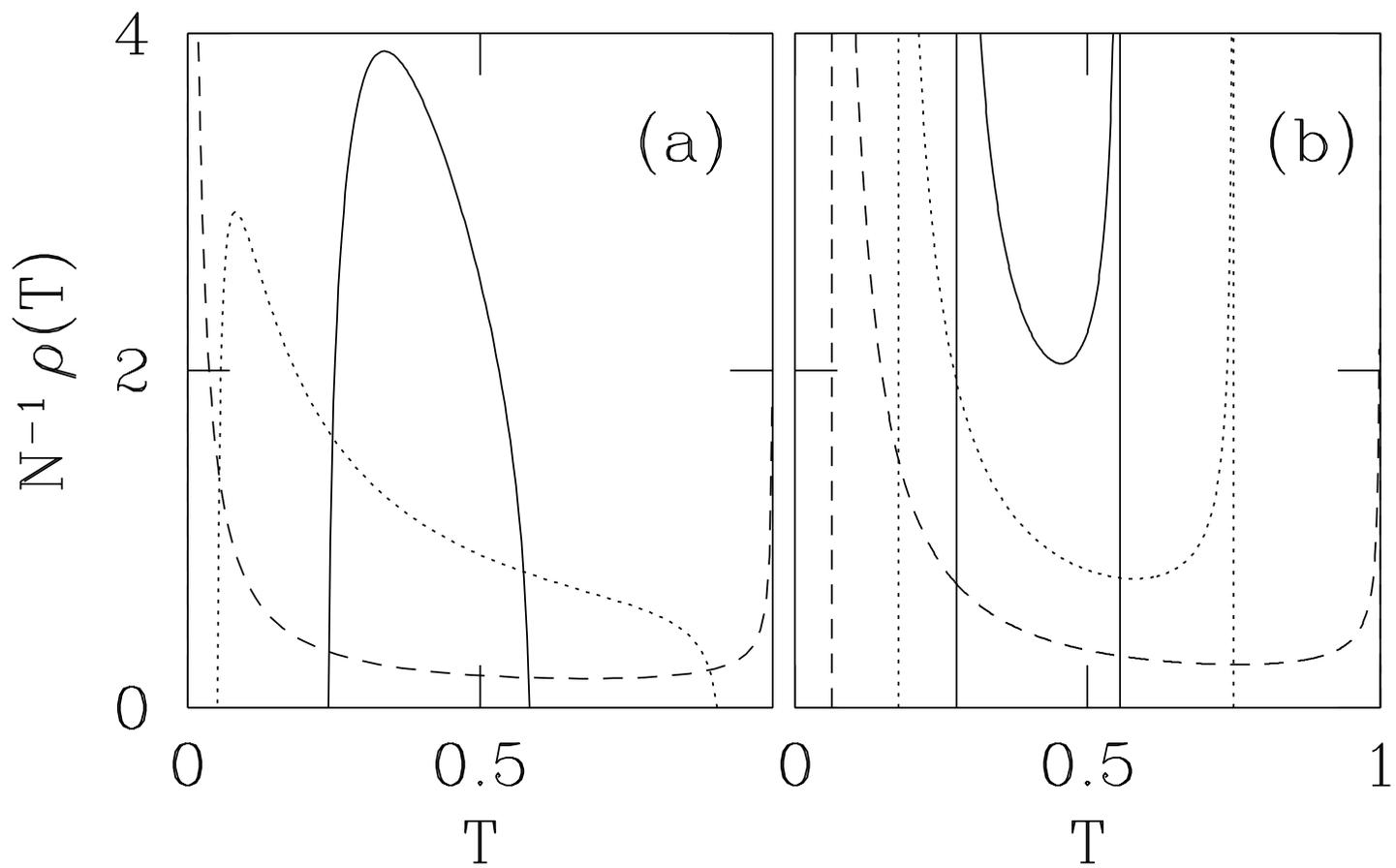
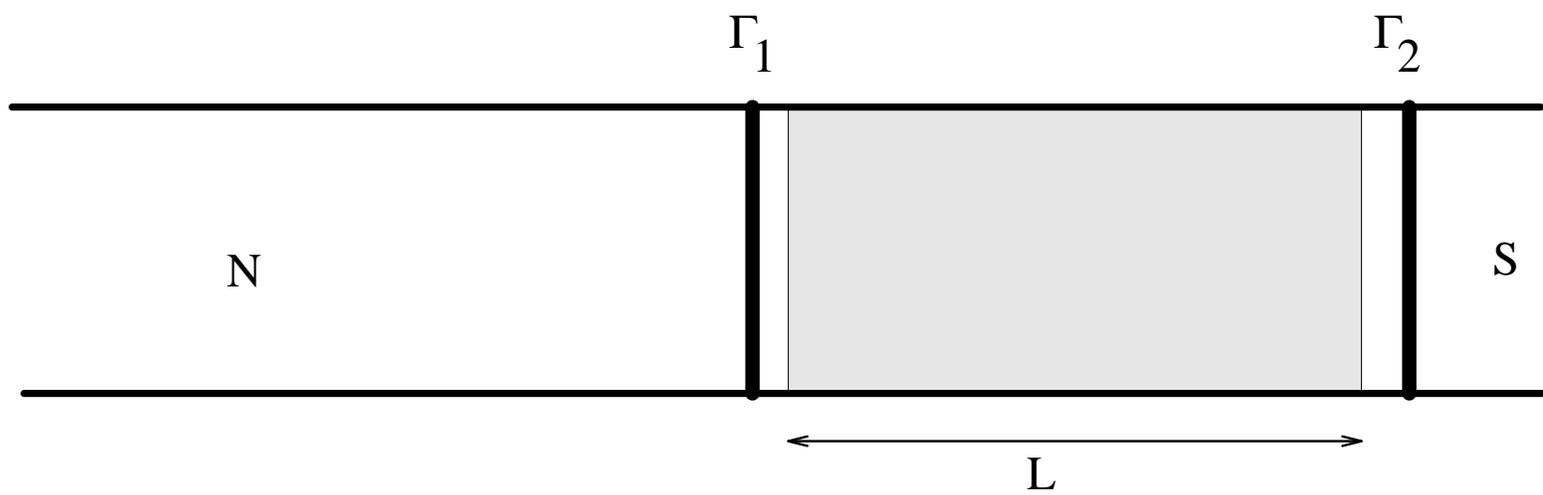
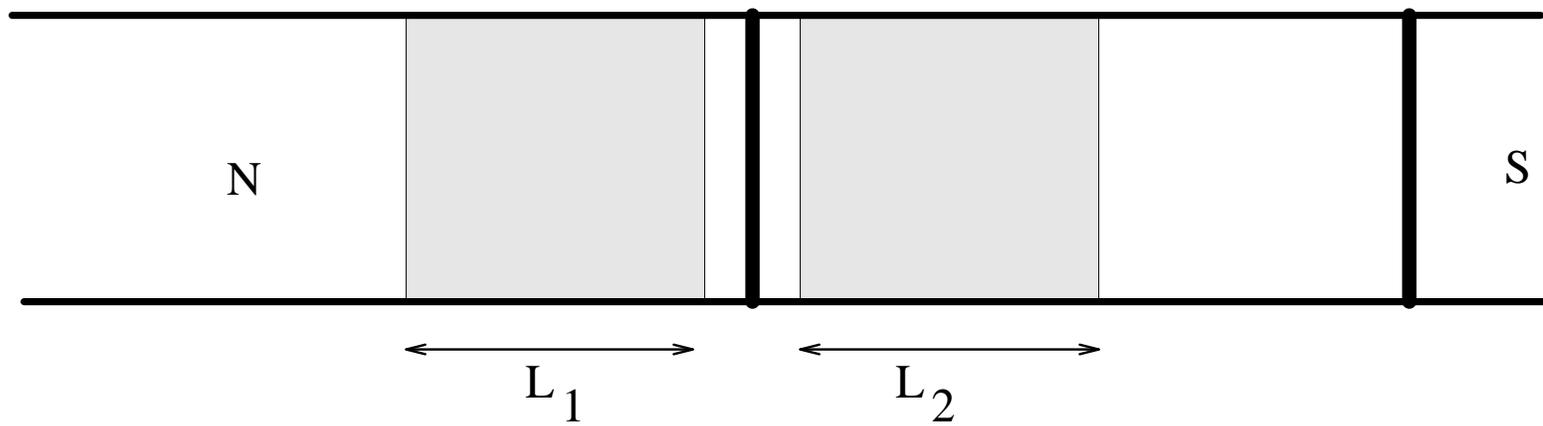
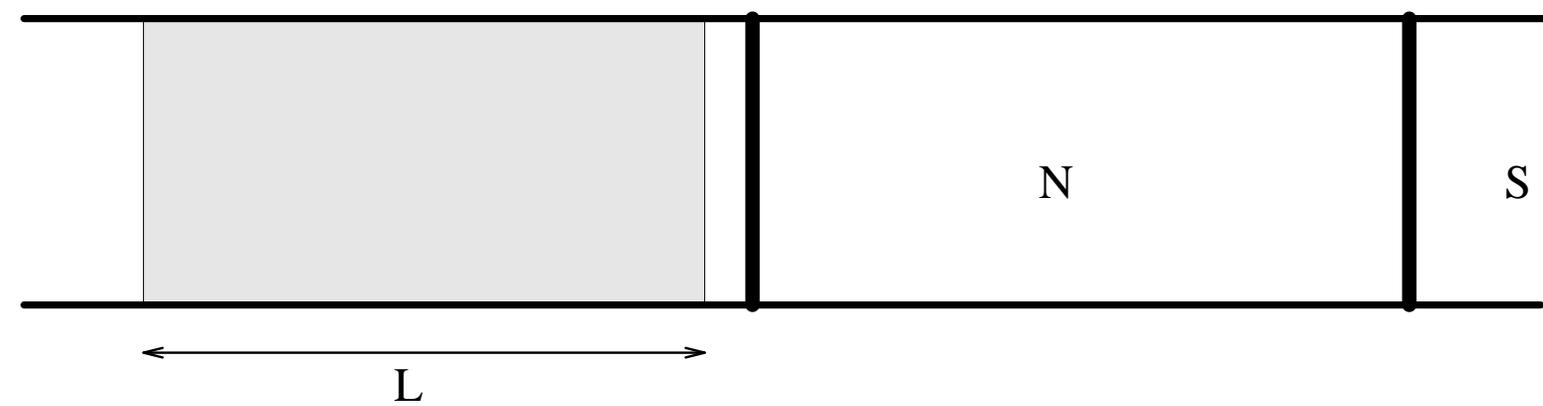
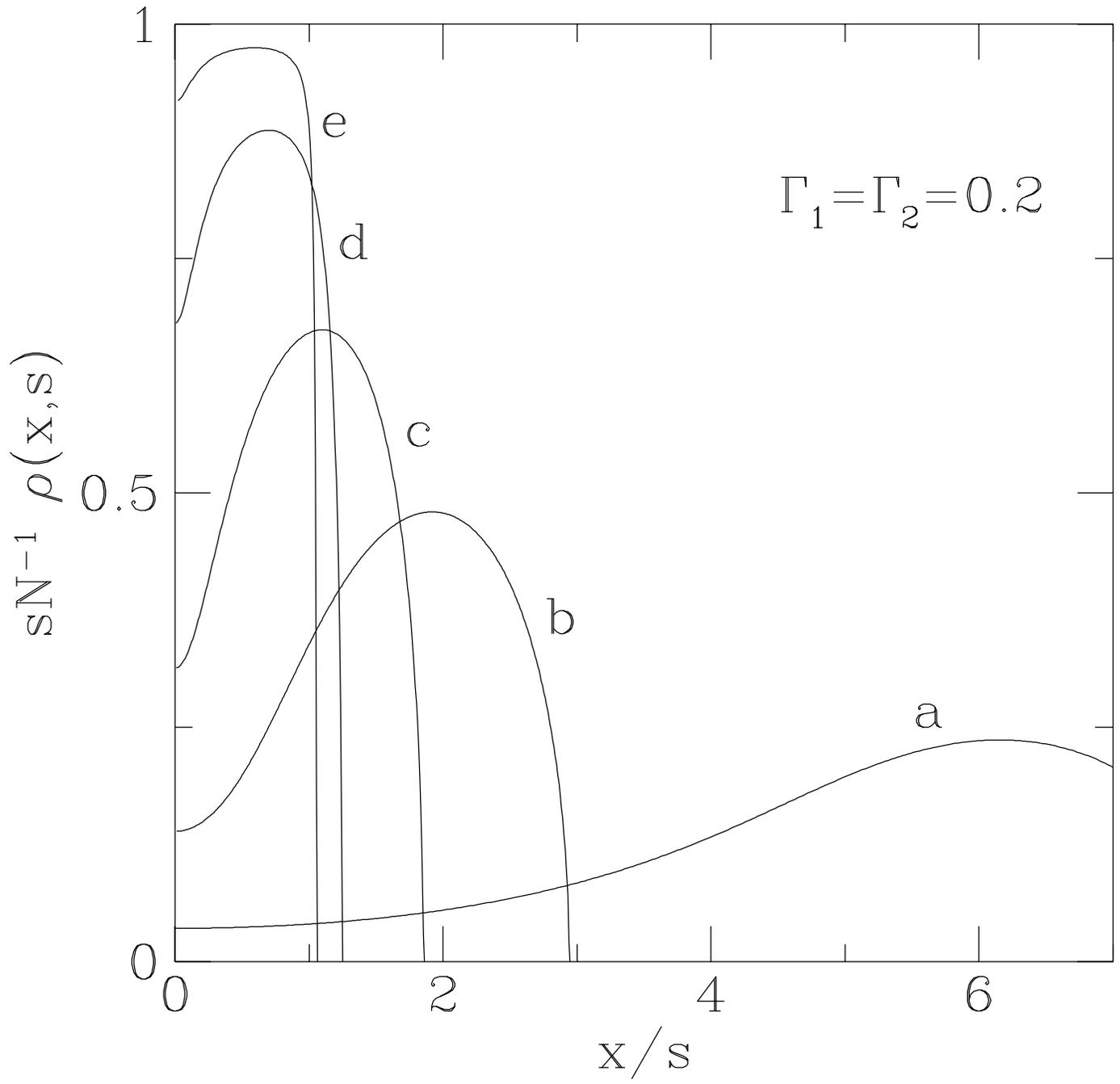
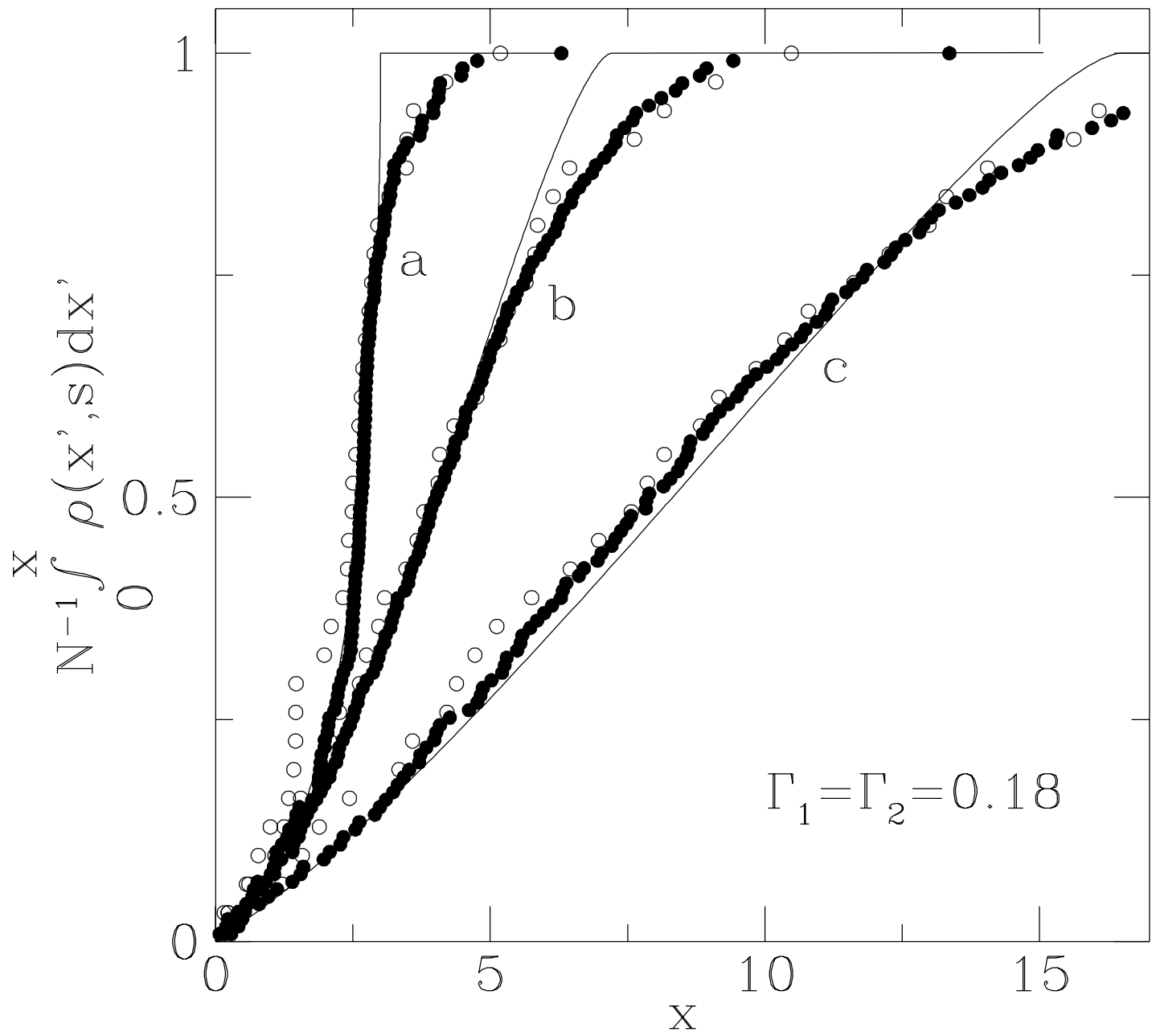
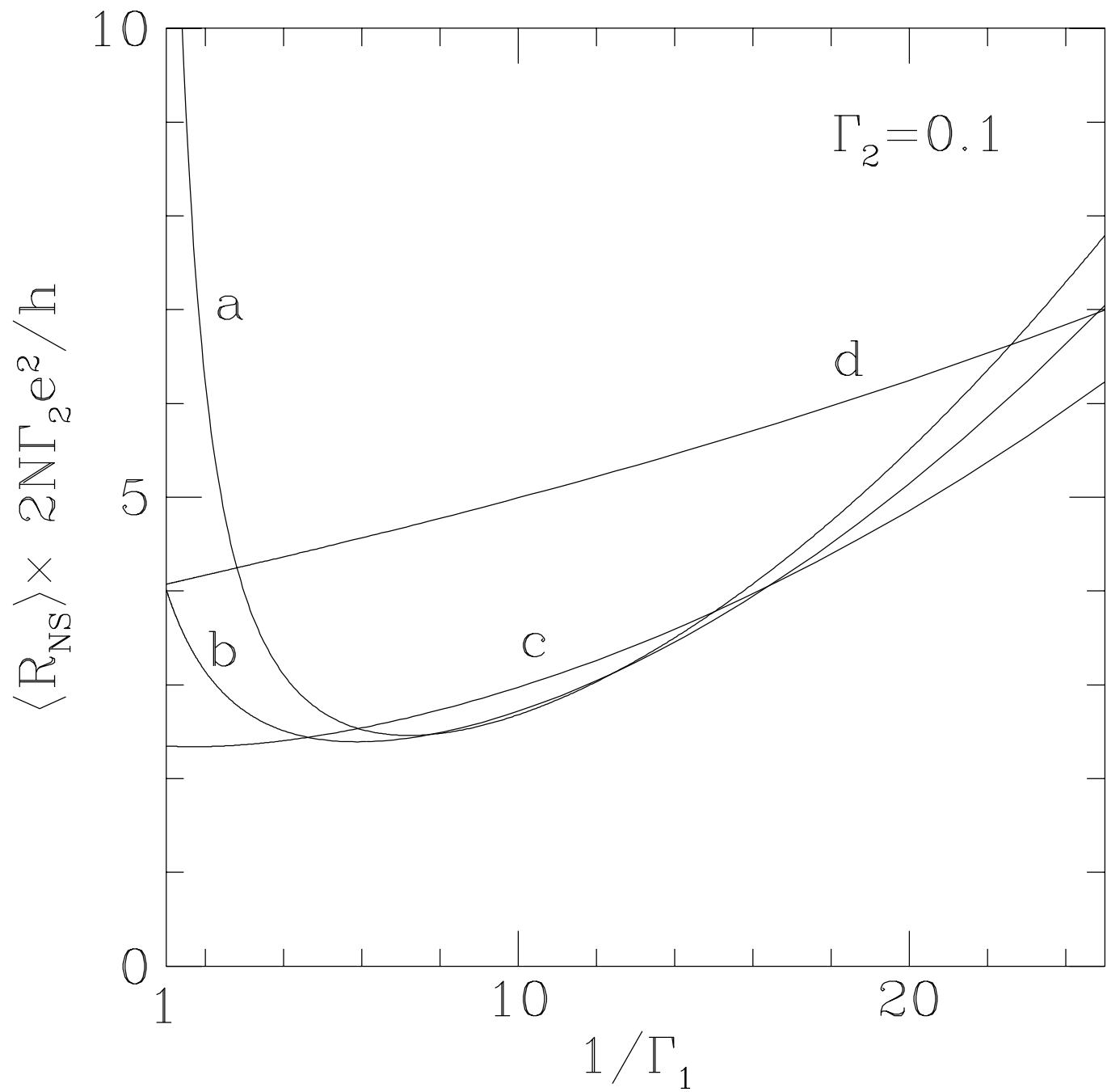
\end{document}